\documentclass[aps,prb,longbibliography,reprint,showkeys,superscriptaddress]{revtex4-1}

\usepackage[utf8]{inputenc}
\usepackage{graphicx}
\usepackage{color}
\usepackage{float}
\usepackage{amsmath}
\usepackage{amssymb}
\usepackage{bm}
\usepackage[version=3]{mhchem}
\usepackage[load=addn]{siunitx}
\usepackage{hyperref}

\newcommand{\vb}[1]{\mathbf{#1}}

\begin{document}
\title{Low Temperature Growth of Graphene on Semiconductor}

\author{H\r{a}kon I. R\o{st}}
\email{hakon.i.rost@ntnu.no}
\author{Rajesh K. Chellappan}
\email{rajesh.k.chellappan@ntnu.no}
\author{Frode S. Strand}
\affiliation{Center for Quantum Spintronics, Department of Physics, Norwegian University of Science and Technology (NTNU), NO-7491 Trondheim, Norway.}
\author{Antonija Grubi\v{s}i\'{c}-\v{C}abo}
\affiliation{School of Physics \& Astronomy, Monash University, Clayton, Victoria 3168, Australia.}
\author{Benjamen P. Reed}
\altaffiliation[Present address: ]{National Physical Laboratory (NPL), Hampton Road, Teddington, TW11 0LW, UK.}
\affiliation{Department of Physics, Aberystwyth University, Aberystwyth SY23 3BZ, United Kingdom}
\author{Mauricio J. Prieto}
\author{Liviu C. T\v{a}nase}
\author{Lucas de Souza Caldas}
\affiliation{Department of Interface Science, Fritz-Haber-Institute of the Max-Planck Society, Faradayweg 4-6, 14195, Berlin, Germany}
\author{Thipusa Wongpinij}
\author{Chanan Euaruksakul}
\affiliation{Synchrotron Light Research Institute, 111 University Avenue, Muang District, Nakhon Ratchasima 30000, Thailand.}
\author{Thomas Schmidt}
\affiliation{Department of Interface Science, Fritz-Haber-Institute of the Max-Planck Society, Faradayweg 4-6, 14195, Berlin, Germany}
\author{Anton Tadich}
\author{Bruce C.C. Cowie}
\affiliation{Australian Synchrotron, 800 Blackburn Rd., Clayton, Victoria 3168, Australia.}
\author{Zheshen Li}
\affiliation{Department of Physics and Astronomy, Aarhus University, Ny Munkegade 120, 8000 Aarhus C, Denmark.}
\author{Simon P. Cooil}
\affiliation{Department of Physics, Aberystwyth University, Aberystwyth SY23 3BZ, United Kingdom}
\affiliation{Semiconductor Physics, Department of Physics, University of Oslo (UiO), NO-0371 Oslo, Norway}
\author{Justin W. Wells}
\email{justin.wells@ntnu.no}
\affiliation{Center for Quantum Spintronics, Department of Physics, Norwegian University of Science and Technology (NTNU), NO-7491 Trondheim, Norway.}

\date{\today}
\begin{abstract}
The industrial realization of graphene has so far been limited by challenges related to the quality, reproducibility, and high process temperatures required to manufacture graphene on suitable substrates. We demonstrate that epitaxial graphene can be grown on transition metal treated 6H-SiC(0001) surfaces, with an onset of graphitization starting around 450-\SI{500}{\celsius}. From the chemical reaction between SiC and thin films of Fe or Ru, $\text{sp}^{3}$ carbon is liberated from the SiC crystal and converted to $\text{sp}^{2}$ carbon at the surface. The quality of the graphene is demonstrated using angle-resolved photoemission spectroscopy and low-energy electron diffraction. Furthermore, the orientation and placement of the graphene layers relative to the SiC substrate is verified using angle-resolved absorption spectroscopy and energy-dependent photoelectron spectroscopy, respectively. With subsequent thermal treatments to higher temperatures, a steerable diffusion of the metal layers into the bulk SiC is achieved. The result is graphene supported on magnetic silicide or optionally, directly on semiconductor, at temperatures ideal for further large-scale processing into graphene based device structures.
\end{abstract}
\maketitle

\section*{Introduction}
Since its experimental discovery in 2004\cite{novoselov2004electric}, graphene - a two-dimensional carbon crystal in a honeycomb structure - has been deemed a promising candidate for device applications due to its exceptional electronic, thermal, optical and mechanical properties\cite{bolotin2008ultrahigh,neto2009electronic,balandin2011thermal,falkovsky2008optical,papageorgiou2017mechanical}. However, the challenges associated with the production of large-scale high-quality graphene layers directly on semiconductor substrates have limited the integration of graphene with conventional device prototypes. 

Until now, the most common techniques for preparing monolayer graphene include micromechanical exfoliation from bulk graphite; epitaxial growth on various transition metals\cite{grapheneNi,grapheneIr,graphenePt,GrapheneRu} through chemical vapour deposition (CVD) of hydrocarbons; and thermal decomposition of bulk crystals such as silicon carbide\cite{berger2004ultrathin,hass2006highly}. Amongst these methods, epitaxial growth by CVD and thermal decomposition of SiC are normally favored as large-area single-crystalline graphene domains can be achieved routinely\cite{coraux2008structural,emtsev2008interaction,emtsev2009towards}. However, CVD grown graphene requires an additional transfer step onto a suitable substrate, limiting the scalability of the technique when it comes to producing graphene on a semiconductor or dielectric of uniform size and quality. The transfer may also introduce contaminants and affect the quality of the CVD graphene, compromising its suitability for device integration\cite{zhang2017rosin}. In comparison, graphene prepared on SiC can be directly converted into a device\cite{lin2010100}, but the temperatures needed to trigger the thermal decomposition of the SiC are by far incommensurate with those of device industry standards\cite{aliofkhazraei2016graphene}. 

In this study, we demonstrate how catalytic reactions between SiC surfaces and thin films of transition metals Fe and Ru can produce quasi-freestanding graphene layers at significantly lower temperatures than those of “conventional” epitaxial growth. Surface graphene layers are formed by allowing the thermally activated metal films to convert $\text{sp}^3$ carbon from the substrate into $\text{sp}^2$ carbon, which reforms at the surface. A similar method has previously been successfully demonstrated at temperatures 500-$\SI{600}{\celsius}$ using Fe on both SiC and diamond\cite{cooil2012iron,cooil2015controlling}. Here, we show that ordered graphene layers can be produced from SiC using either Fe or Ru, with an onset of growth starting at around 450-\SI{500}{\celsius}. This metal-mediated approach leaves graphene resting on underlying layers of metal silicide which can then be eliminated by subsequent thermal treatments to higher temperatures, driving diffusion of the metal ions into the bulk crystal\cite{shen2018fabricating}. The result is quasi-freestanding graphene resting on semiconducting substrates and grown at industrially compatible temperatures, ideal for further processing into large-scale, graphene-based device structures.  

\begin{figure*}[t]
  \centering
  \includegraphics[width=0.9\textwidth]{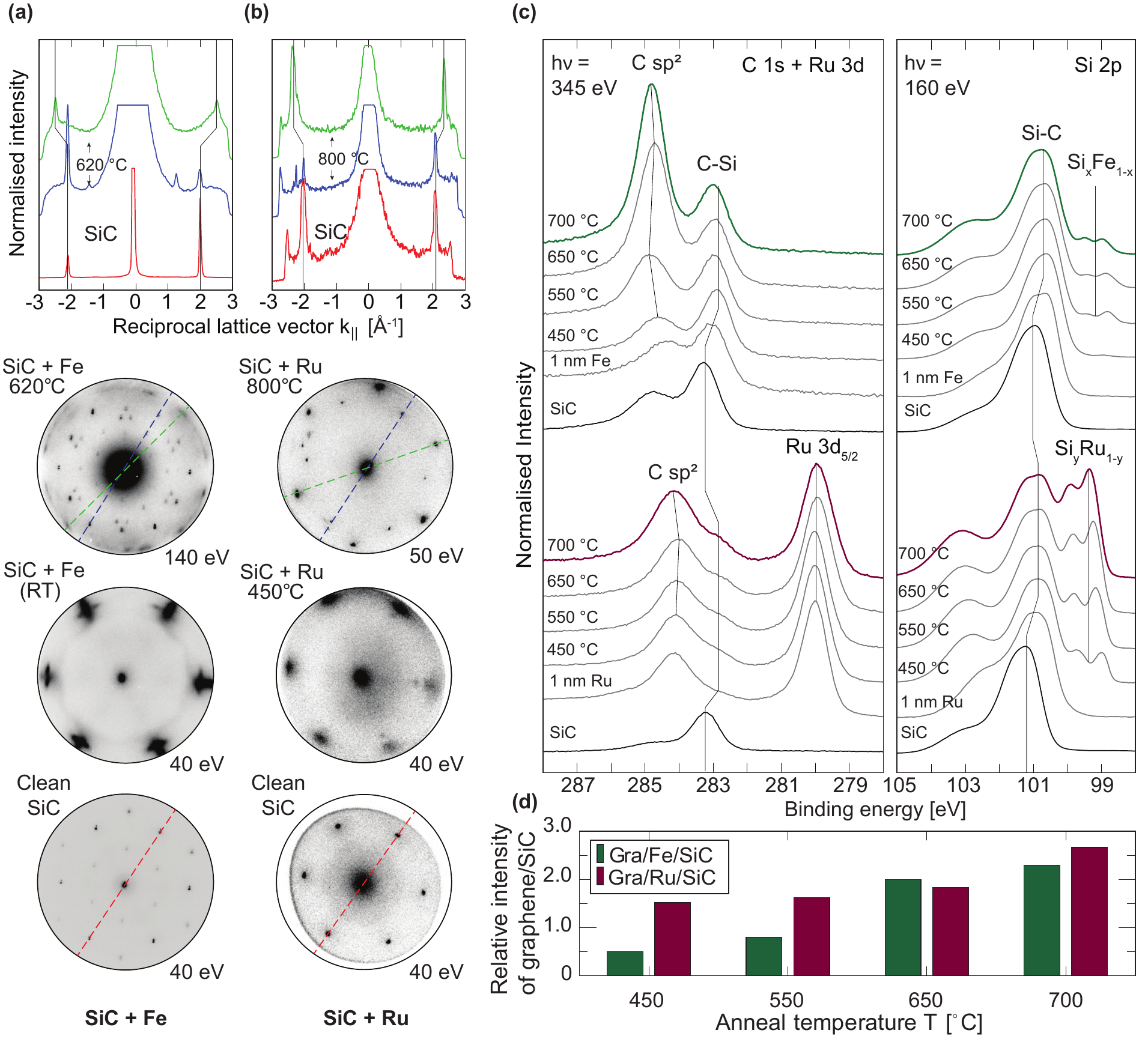}
    \caption{\textbf{Gradual formation of graphene on SiC treated with Fe and Ru.} \textbf{(a)} and \textbf{(b)}: $\mu-$LEED patterns of 6H-SiC coated with thin films of Fe and Ru, respectively, and then thermally treated to temperatures 600-\SI{800}{\celsius}. The excitation energy/starting voltages were adjusted for the final images to increase the brightness of all relevant spots. Intensity cuts along high-symmetry directions of the reciprocal lattices of the two systems (topmost panels) reveal new diffraction maxima appearing with the formation of graphitic carbon at the surface, at angles $30^\circ$ (SiC + Ru), $15^\circ$ and $45^\circ$ (SiC + Fe) relative to the underlying SiC. All intensity cuts were normalised to the $\vb{k}$ values and intensities of the 1st order SiC spots.  (\textbf{c}) \ce{Si} 2p, \ce{C} 1s and \ce{Ru} 3d signal showing the formation of graphitic layers: silicon carbide, followed by deposition of \ce{Fe} and \ce{Ru} and successive thermal treatments in the range $450-$\SI{700}{\celsius}. (\textbf{d}) The estimated ratio of of graphitic carbon to silicon carbide signal measured within a few nanometers of the surface, as a function of temperature.}
    \label{fig:figure1}
\end{figure*}

\section*{Results and Discussion}
The gradual formation of graphitic carbon at the surface of Fe/SiC and Ru/SiC as a function of temperature is illustrated in FIG.\ \ref{fig:figure1}. Chemically and thermally cleaned SiC crystals treated with thin films of Fe and Ru were studied after metallization and subsequent annealing to higher temperature using small-spot low-energy electron diffraction ($\mu$-LEED) and high resolution X-ray photoelectron spectroscopy (XPS). 

FIG.\ \ref{fig:figure1}a and \ref{fig:figure1}b show the development of the surface diffraction patterns of the two systems. Following high temperature cleaning, both samples initially display known surface reconstructions of the SiC(0001) face: namely a ($1\times1$) phase and a $(\sqrt{3}\times\sqrt{3})\text{R}30^\circ$ silicate reconstruction, both indicating a clean surface\cite{starke1999sic}. With metallization, the additional Fe and Ru layers both seem to mimic the hexagonal arrangement of the Si atoms with a slight lattice mismatch: the newfound diffraction spots both appear at larger magnitudes of the momentum wave vector  $\vb{k}$ than the first order SiC spots. Upon thermal activation, the metallic spots disappear and the original ($1\times1$) phase of the SiC(0001) face again becomes visible, as seen from comparing the intensity profiles along the same high symmetry direction for the clean SiC (red dashed line) and after the thermal treatments (blue dashed line). The recurrence of the SiC pattern is also accompanied with several new diffractive features, some of which are similar to previous reports of the Fe/SiC system\cite{cooil2012iron}. Notably, both systems show higher $\vb{k}$ features appearing at rotations $15^\circ$ and $45^\circ$ (Fe), or $30^\circ$ (Ru) relative to the SiC spots. Investigating these more closely along their symmetry directions (green dashed line) and comparing them to the simultaneously visible SiC features reveal that the new spots occur at $\lvert\vb{k}\rvert\sim\SI{2.36}{\angstrom}^{-1}$ for Ru/SiC and $\lvert\vb{k}\rvert\sim\SI{2.50}{\angstrom}^{-1}$ for Fe/SiC. Both values are within $10\%$ of the $\lvert\vb{k}\rvert$ value associated with pristine graphene flakes of lattice constant $a=\SI{2.46}{\angstrom}$\cite{ohta2006controlling}. At first glance, the newfound spots on Ru/SiC could be interpreted to come from the SiC surface, as their $\lvert\vb{k}\rvert$ value is similar to the second order spots of the $(\sqrt{3}\times\sqrt{3})\text{R}30^\circ$ silicate reconstruction. However, no first order spots from this reconstruction are observed and hence the $\lvert\vb{k}\rvert\sim\SI{2.36}{\angstrom}^{-1}$ spots cannot be explained from the SiC surface alone.

FIGs.\ \ref{fig:figure1}c and \ref{fig:figure1}d show the C~1s and Si~2p core levels for both material systems on samples with similar preparation. Both regions were acquired from photoelectrons with kinetic energies corresponding to a shallow escape depth of $\lambda\sim\SI{0.5}{\nm}$ beneath the sample surface. Similar features indicative of clean SiC can be seen for both systems: dominant peak components from the Si-C bonding of the substrate can be found at binding energies \SI{283.2}{\eV} (C~1s) and \SI{101.2}{\eV} (Si~2p), with minor features at $\sim$\SI{284.8}{\eV} and \SI{103.5}{\eV}, indicating some occurrence of C-C bonding\cite{nakao1992xps,wang2012xps} and silicon oxide formation\cite{hollinger1984probing,watanabe2011synchrotron}. With metallization, a shift of $\sim$\SI{0.4}{\eV} can be seen in the bulk components, indicating a charge transfer to the surface SiC from the overlaying metal. 

Annealing the samples to \SI{450}{\celsius} and then \SI{550}{\celsius} reveals new components in the Si~2p regions at binding energies 98-$\SI{100.5}{\eV}$, indicative of a reaction between the metallic thin films and the underlying Si-rich layers. Similar features have been reported previously and associated with the formation of transition metal silicide phases\cite{cooil2012iron,ruhrnschopf1996growth,gomoyunova2007initial,wang2014graphene,lu1991photoemission,lizzit2012transfer}. The occurrence of new chemical species is further supported by the observed change of intensity between the components in the C~1s region, and the broadening of the transition metal core levels. An increased signal from the higher binding energy components of C~1s suggests that new species of carbon have been formed, consistent with the previously reported transition metal-mediated liberation of C atoms from SiC that reform into graphitic carbon\cite{cooil2012iron}. 

Further thermal treatments to \SI{650}{\celsius} and \SI{700}{\celsius} reveal a continuation of the same trend, where the graphitic carbon and silicide components continue to grow while signal is lost from the transition metal and bulk SiC components. To quantify the amount of graphitic carbon formed from the reaction, the C~1s region was deconvolved into peak components corresponding to C $\text{sp}^2$ and C-Si signal. For the Ru/SiC system, the Ru $3\text{d}_{5/2}$ was initially fitted and used to estimate and deduct the intensity of Ru $3\text{d}_{3/2}$ features at the appropriate binding energy in the region overlapping with the C~1s components (this will be discussed later). FIG.\ \ref{fig:figure1}d shows the development of the (C $\text{sp}^{2}$)/(C-Si) signal ratio with temperature: at \SI{700}{\celsius} the C $\text{sp}^{2}$ signal is more than double that of the bulk. Accompanied with the attenuation of the metal signal, this suggests that the higher temperatures trigger graphitic carbon to form near the sample surface\cite{cooil2012iron,cooil2015controlling}. 

\begin{figure*}[t]
\centering
    \includegraphics[width=\textwidth]{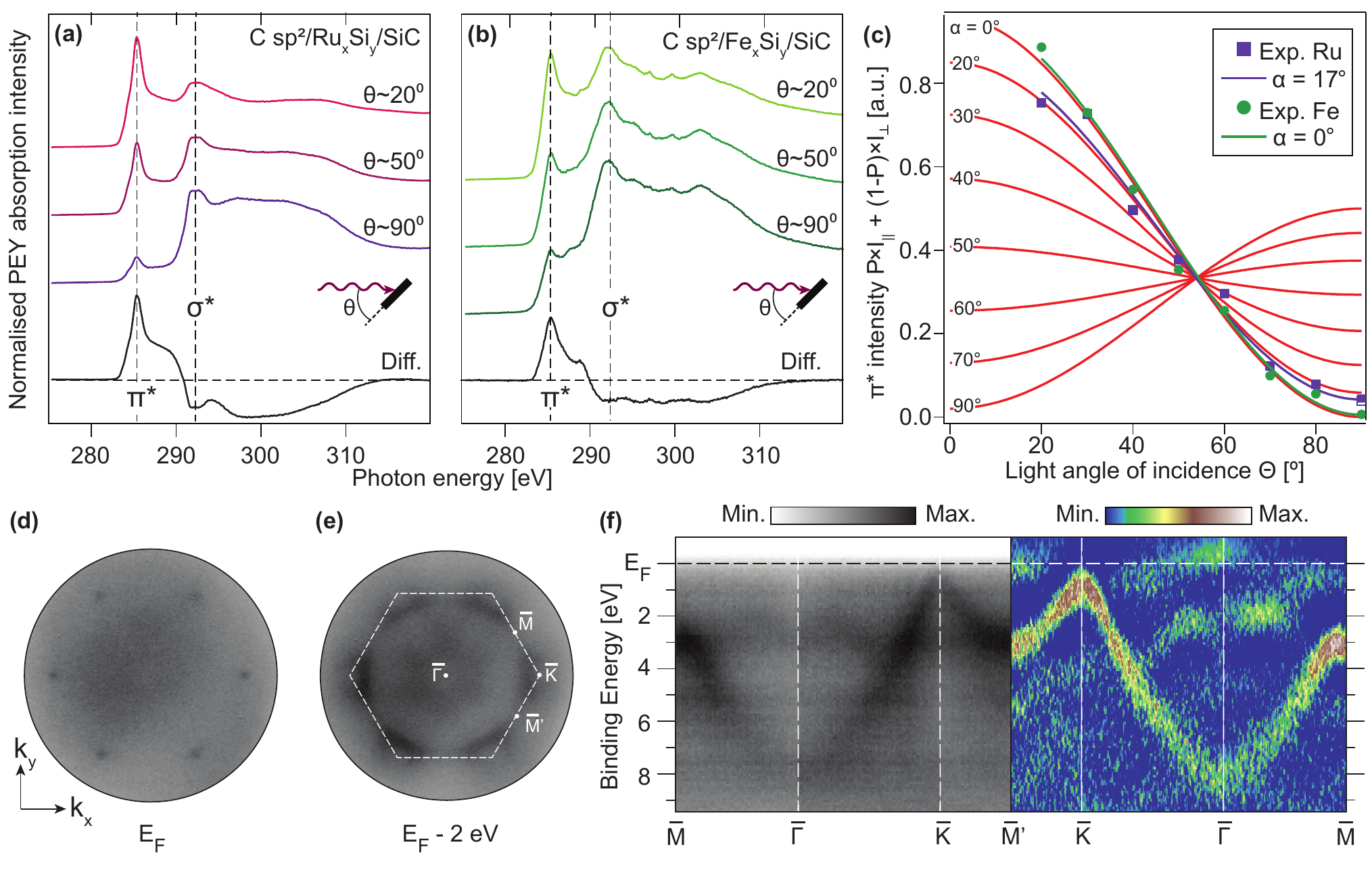}
    \caption{\textbf{C~1s absorption features, layer orientation and electronic structure of surface graphene. }\\\textbf{(a)-(b)} Angle-dependent NEXAFS for graphene on \ce{Ru}/\ce{SiC} (a) and \ce{Fe}/\ce{SiC} (b) ranging from roughly grazing $\left(\sim\ang{20}\right)$ to normal incidence. Prominent resonance features from the \ce{C} 1s $\pi^{*}$ and $\sigma^{*}$ transitions can be seen at excitation energies \SI{285.5}{\eV} and \SI{291.7}{\eV}, respectively. A difference trace between the grazing and normal incidence spectra is also shown for each system to highlight the intensity changes with the light orientation. \textbf{(c)} Intensities of the LUMO (1s$\rightarrow\pi^{*}$) absorption of graphene grown on \ce{Ru}/\ce{SiC} (purple) and \ce{Fe}/\ce{SiC} (green), plotted against the theoretical curves $I\left(\theta,\alpha\right)$ as described in Eqn.~\ref{eq:NEXAFSint}. A best-fit to each data set gives an average angle $\alpha<20^\circ$ between the substrate surface normal and the vector perpendicular to the graphene layers. \textbf{(d)-(e)} Constant $E$ vs. ($\vb{k}_{x}$,$\vb{k}_{y}$) maps of the gra/\ce{Fe}/\ce{SiC} system measured at $E_{\text{F}}$ (d) and $E_{\text{F}}-\SI{2}{\eV}$ (e) with an excitation energy of $h\nu=\SI{115}{\eV}$. The first Brillouin zone boundary (dashed line) and high symmetry points $\Bar{\Gamma}$, $\Bar{\text{M}}$, $\Bar{\text{K}}$ and $\Bar{\text{M}}'$ have been marked. (\textbf{f}) $E$ vs. $\vb{k}_{||}$ intensity plots of occupied states of the gra/\ce{Fe}/\ce{SiC} system along principal directions $\Bar{\text{M}}\rightarrow\Bar{\Gamma}$, $\Bar{\Gamma}\rightarrow\Bar{\text{K}}$ and $\Bar{\text{K}}\rightarrow\Bar{\text{M}}'$. The plots to the left show the background corrected bandstructure, while on the right a second derivative image (SDI) with a moderate boxcar averaging has been applied to amplify the graphene energy dispersion relative to the background.}
    \label{fig:figure2}
\end{figure*}

FIG.\ \ref{fig:figure2}a and \ref{fig:figure2}b show the near edge X-ray absorption fine structure (NEXAFS) of the C~1s K-edge from the Ru/SiC and Fe/SiC systems, respectively. The spectra were recorded from samples similar to those in FIG.\ \ref{fig:figure1}, before and after thermal treatments to \SI{800}{\celsius} using linearly polarized light at angles ranging from grazing ($\sim20^\circ$) to normal incidence relative to the 6H-SiC(0001) plane. In both cases the annealing leads to new absorption resonances appearing at excitation energies \SI{285.5}{\eV} and \SI{291.7}{\eV}. While the latter indicates the existence of antibonding orbitals which overlap head-on ($\sigma^{*}$), the former is a fingerprint of antibonding $\text{p}_{z}$ orbitals in $\text{sp}^{2}$ hybridized carbon, perpendicularly oriented to an aromatically configured macromolecular plane\cite{pacile2008near,shpilman2014near}. The strong $1\text{s}\rightarrow\pi^{*}$ resonance therefore supports the formation of new graphitic carbon on the samples. 

The $\pi^{*}$ resonances in FIG.\ \ref{fig:figure2} also reveal a definite angle dependence, with stronger intensities observed when electrons are exited by light near grazing incidence to the samples. This suggests a prominent geometric ordering in the layers\cite{kim2008temperature}. In order to determine the orientation of the C-C bonds relative to the plane of the SiC substrate, we compared the intensity of the lowest unoccupied molecular orbital (LUMO) excitation to an analytical solution of the NEXAFS intensity outlined by St\"{o}hr and Outka\cite{stohr1987determination,stohr2013NEXAFS}. Essentially, for carbon systems of threefold or higher symmetries, the resonance intensity of the $1\text{s}\rightarrow\pi^{*}$ transition when excited using linearly polarized light can be reduced to:
\begin{align}\label{eq:NEXAFSint}
    I \propto\quad &P\left(\cos^{2}\theta\cos^{2}\alpha + \frac{1}{2}\sin^{2}\theta\sin^{2}\alpha\right)\nonumber \\
    &+ \frac{(1-P)}{2}\sin^{2}\alpha,   
\end{align}
where $P$ is the polarization factor of the light hitting the sample plane at angle $\theta$. The angle $\alpha$ between the sample surface normal and the vector perpendicular to the molecular plane of the inherent C-C bonds can thus be used to determine the average orientation of the graphitic planes on the surface.

In FIG.\ \ref{fig:figure2}c, the recorded LUMO intensities for the carbon in both material systems are plotted against the theoretical curves given by Eqn.~\ref{eq:NEXAFSint} for light with nearly perfect linear polarization ($P>0.9$). In order to establish the angular intensity from the newly formed C-C carbon alone, a Gaussian profile was fitted to the $\pi^*$ absorption feature for incidence angles $\theta=20-90^\circ$, and the background intensity of the clean SiC surface was subsequently deducted from the region. For the Ru/SiC system an additional, but presumably not angularly dependent, constant intensity offset was added to the Gaussians to account for any excitations from the Ru 3d state. Both samples show an average tilt angle $\alpha<20^\circ$ suggesting that the graphitic layers are oriented more or less parallel to the underlying substrate.

Properties related to the electronic structure of the newfound carbon layers, such as the number of layers formed, can be deduced from an ARPES map of the occupied electronic states of the system\cite{ohta2006controlling,bostwick2007quasiparticle,ohta2007interlayer,siegel2010quasifreestanding}. FIGs.\ \ref{fig:figure2}d and \ref{fig:figure2}e show the occupied electronic states of the Fe/SiC system after annealing to \SI{600}{\celsius}, pictured in reciprocal space ($\vb{k}_{\text{x}}$ vs. $\vb{k}_{\text{y}}$) at constant energies relative to the Fermi level $(E_{\text{F}})$. Near $E_{\text{F}}$, six distinguishable features from the $\pi$ bands, i.e. Dirac ``cones'', appear at the $\bar{\text{K}}$ and $\bar{\text{K}}'$ points of the Brillouin zone. Roughly \SI{2}{\eV} below $E_{\text{F}}$, another broad, hexagonal feature of smaller $\vb{k}$ values and with a $30^\circ$ relative rotation can be seen. In FIG.\ \ref{fig:figure2}f, intensity plots of the occupied states have been extracted for $E$ vs. $\vb{k}_{||}$ along the principal symmetry directions of the surface Brillouin zone. Two similar plots are shown, where one (left) is the photoemission signal extracted from the constant energy surfaces and the other (right) is a second differential image (SDI) of the same plot with a moderate boxcar averaging applied. In both plots, the characteristic, near linearly dispersive bands of graphene are present close to the $\bar{\text{K}}$ point, indicating that at least one surface layer of graphene has been formed from the reaction. The SDI also reveals the additional, low $\vb{k}$ dispersive features around the $\Bar{\Gamma}$ point at roughly \SI{2}{\eV} binding energy. The exact origin of these bands is uncertain, but assumed to arise from the metal's shallow d states and their interaction with the SiC surface.

\begin{figure}
    \centering
    \includegraphics[width=0.9\linewidth]{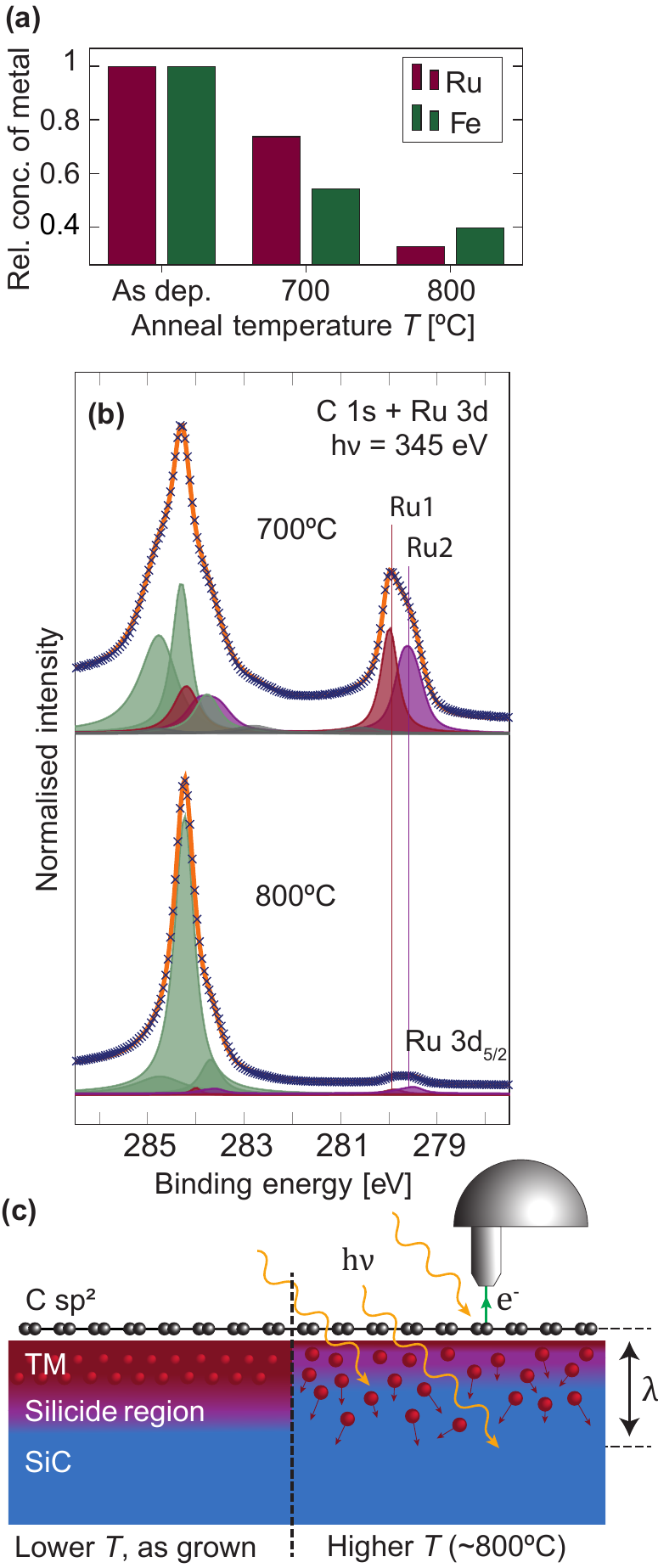}
    \caption{\textbf{Diffusion of metal into the growth substrate.} \textbf{(a)} The relative concentration of metal in the surface layers before and after the heat treatments. The measure of the metal content is given as the ratio of \ce{Ru} and \ce{Fe} core level signal to the substrate carbon signal, corrected for differences in cross-sections and normalised to unity after deposition (room temp.). Continued diffusion into the ``bulk'' \ce{SiC} is evident for subsequent heat treatments to higher temperatures. \textbf{(b)} \ce{C} 1s and \ce{Ru} 3d core levels of \ce{Ru} on silicon carbide, measured with photon energy $h\nu=\SI{345}{\eV}$ after subsequent heat treatments to \SI{700}{\celsius} and then \SI{800}{\celsius}. \textbf{(a)} Schematic of metal diffusion into the bulk of the \ce{SiC} substrate during annealing to \SI{800}{\celsius}.}
    \label{fig:figure3}
\end{figure}

With subsequent annealing to higher temperatures, both the gra/Fe/SiC and gra/Ru/SiC system experience further changes to their chemical and structural composition (FIG.\ \ref{fig:figure3}). Not surprisingly, the relative intensity of graphene to bulk SiC signal increases (FIG.\ \ref{fig:figure1}d), suggesting more metal has reacted with the SiC, liberating more carbon. Second, the relative intensity of the metal to bulk signal is seen to decrease, suggesting that metal is disappearing from the surface layers. FIG.\ \ref{fig:figure3}a shows the change in concentration of Fe and Ru in the topmost few nanometers of both systems with increasing temperature. After two short heat treatments to higher temperatures \SI{700}{\celsius} and \SI{800}{\celsius}, the relative concentration of metal to bulk SiC is less than 50\% of the initial value. Similar behavior has been observed  for intercalated ad-atoms of Fe\cite{sung2014spin,shen2018fabricating} and Yb\cite{watcharinyanon2013ytterbium} at temperatures beyond \SI{600}{\celsius}. The attenuation of Fe was explained by a temperature-driven diffusion into the underlying semiconductor substrate and the same argument could easily be extended to Ru.

\begin{figure*}[t] 
\centering
    \includegraphics[width=\textwidth]{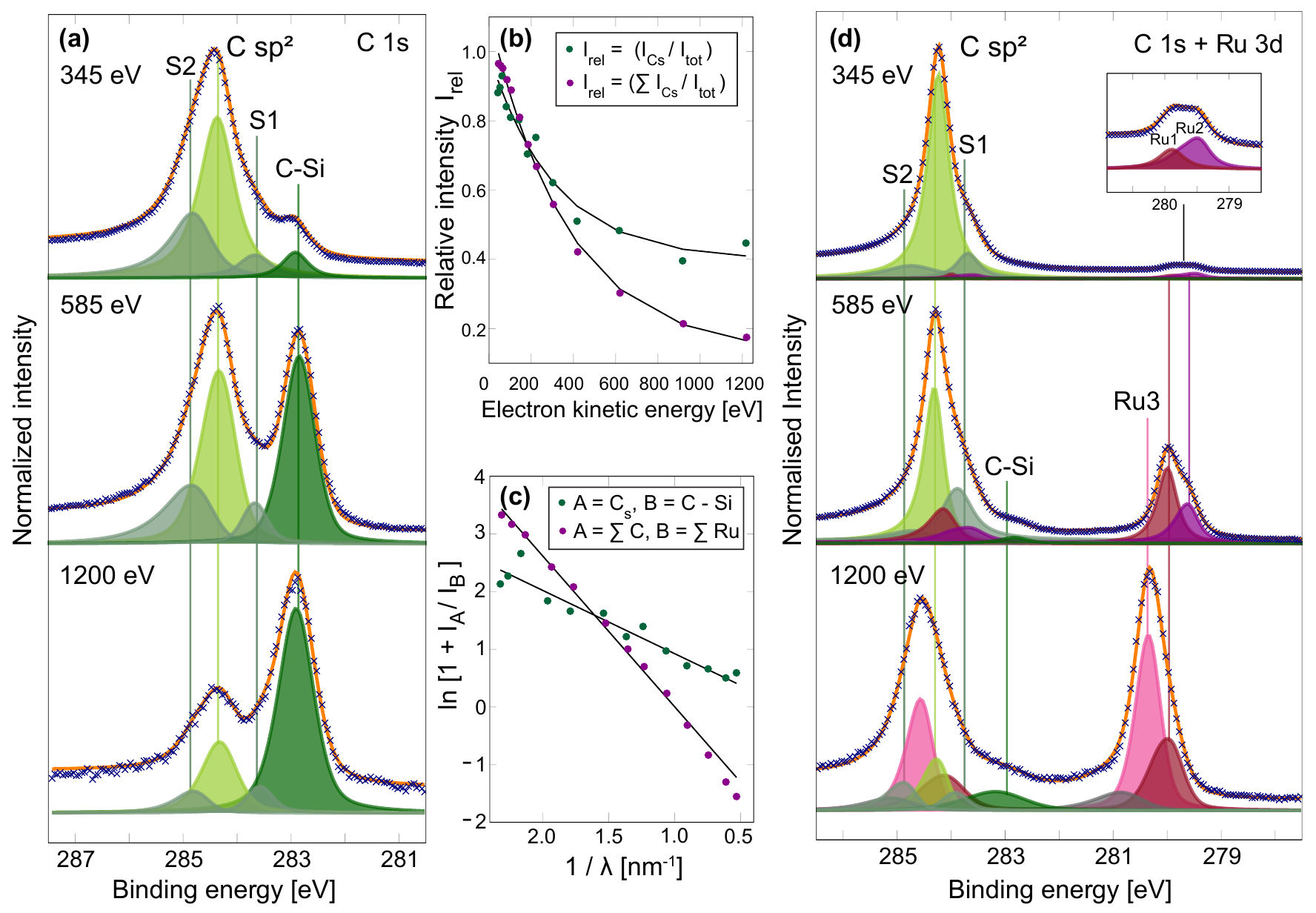}
    \caption{\textbf{Depth profiling of graphene from mediated growth using \ce{Fe} and \ce{Ru} on \ce{SiC}. } \textbf{(a)} and \textbf{(d)}: a selection of \ce{C} 1s core level spectra from samples covered with \ce{Fe} and \ce{Ru}, respectively, after annealing to \SI{800}{\celsius}. The three core levels in each set were measured with photon energies $h\nu=\SI{345}{\eV}$ (surface sensitive), $h\nu=\SI{585}{\eV}$ (moderately surface sensitive) and $h\nu=\SI{1200}{\eV}$ (bulk sensitive). \textbf{(b)}: a full range of data points for the photoelectron intensity of graphene ($I_{\text{Cs}}$) relative to the total intensity ($I_{\text{tot}}$) from the energy regions in (a) and (d), as a function of kinetic energy for the exited photoelectrons. Note that the two data sets differ slightly in their normalisations, with green (gra/Fe/SiC) stating the fractional intensity of \ce{C} $\text{sp}^{2}$ plus surface peaks (S1, S2), while for purple (gra/Ru/SiC) all carbon features were assumed to be surface carbon for simplicity. \textbf{(c)}: the data from (b), re-plotted to show the relationship between the relative photoelectron intensity and the inverse inelastic mean-free path $\lambda$ from a simple two-layer model: surface graphene ($I_{\text{A}}$) on top of an underlying, metal-rich substrate region ($I_{\text{B}}$). An evident linearity can be seen from the data, similar to that of the ideal two-layer model. As seen from (a) and (d) the \ce{Ru} signal, but not \ce{Fe} signal, can be seen in the relevant energy regions of the \ce{C} 1s core level. Hence for the gra/Fe/SiC system, the \ce{C} $\text{sp}^{2}$ intensities (top layer) are compared to the carbon signal from the underlying \ce{SiC} substrate (bottom layer), while for the gra/Ru/SiC system the metallic \ce{Ru} signal is considered to be the bottom layer.}
    \label{fig:figure4}
\end{figure*}

Assuming the metal is depleted from the surface layers by thermal diffusion, one could expect that probing deeper into the samples would reveal a relatively stronger XPS signal from the diffused metallic species. FIG.\ \ref{fig:figure4} shows the C~1s region of the gra/Fe/SiC and gra/Ru/SiC samples discussed in FIG.\ \ref{fig:figure3}, probed with increasingly higher photoexcitation energies. Higher $h\nu$ gives longer inelastic mean-free paths $\lambda$ for the excited photoelectrons, and so signal from an increasingly thick slab of substrate material is being detected as $h\nu$ increases. FIG.\ \ref{fig:figure4}a shows the C~1s region of gra/Fe/SiC, recorded with three selected photoexcitation energies corresponding to $\lambda\sim\SI{0.5}{\nm}$ (surface sensitive), $\lambda\sim\SI{0.9}{\nm}$ (moderately surface sensitive) and $\lambda\sim\SI{1.6}{\nm}$ (bulk sensitive). The signal can be deconvolved into two main features at \SI{282.9}{\eV} and \SI{284.3}{\eV}, where the former matches the previously observed bulk SiC component, shifted due to interactions with the transition metal. The second feature is found at a binding energy roughly \SI{0.1}{\eV} lower than that expected for neutral graphite, and has a distinct asymmetric shape indicating a metallic surface nature\cite{emtsev2008interaction,emtsev2009towards}. At high surface sensitivity ($\lambda\sim\SI{0.5}{\nm}$), the sharp asymmetric feature dominates the intensity of the region, but at lower surface sensitivity ($\lambda\rightarrow\SI{1.6}{\nm}$) the weight is shifted towards the substrate peak. This matches well with the notion of a finite amount of graphene situated at the surface: at low $\lambda$ most of the C~1s signal comes from the  layers of $\text{sp}^{2}$ carbon, but at higher $\lambda$ more substrate layers are added to the probing region and so the bulk signal dominates.
In FIG.\ \ref{fig:figure4}d, the same exercise is repeated for the gra/Ru/SiC system using the same photoexcitation energies. Note that the deconvolution is more complicated due to the overlapping C~1s and Ru 3d signal. However, from the known energy separation and intensity ratio of the $3\text{d}_{5/2}$ and $3\text{d}_{3/2}$ signals, one can accurately determine the components of Ru $3\text{d}_{3/2}$ based on those observed in the Ru $3\text{d}_{5/2}$ region. The overlapping C~1s and Ru $3\text{d}_{3/2}$ signal was thus deconvolved by placing features of the bare SiC and the pre-determined Ru components together, and adding the minimum number of additional features to optimize the fit.

Not surprisingly, the same distinct asymmetric feature seen in FIG.\ \ref{fig:figure4}a is also visible for gra/Ru/SiC in FIG.\ \ref{fig:figure4}d, now at \SI{284.2}{\eV}. Near the surface ($\lambda\sim\SI{0.5}{\nm}$) the bulk SiC feature cannot be seen, but with increasing $\lambda$ it again becomes visible at lower binding energy \SI{282.8}{\eV} as was observed in the gra/Fe/SiC system. At moderate depth sensitivity, the Ru contributes with two distinct features: one at \SI{280.0}{\eV} (Ru1) corresponding to that of metallic Ru\cite{fuggle1975photoelectron,kaga1999ru}, and one at \SI{0.4}{\eV} lower binding energy (Ru2) matching with that observed for Ru silicides\cite{lu1991photoemission,lizzit2012transfer,pasquali2009formation}. With increasing depth sensitivity, the relative weight of the two changes and new Ru features become visible at higher binding energies (Ru3). This indicates that different Ru species are present at different depths into the substrate, in line with the diffusion of metal into the bulk upon thermal activation beyond \SI{600}{\celsius}.

To correctly determine the position of the graphene layers relative to the metal silicide and the SiC bulk, the intensity of the C~1s signal was extracted for additional photoexcitation energies in the range $315-\SI{1200}{\eV}$. FIG.\ \ref{fig:figure4}b shows how the graphene intensity varies relative to the total accumulated intensity in the scan region for both material systems. In the gra/Fe/SiC system the relative graphene intensity was estimated by first deconvolving the C~1s region, and then comparing the sum of the asymmetric feature at \SI{284.3}{\eV} and the surface peaks to the total signal. In the case of gra/Ru/SiC, the graphene signal was estimated by first determining the total Ru $3\text{d}_{5/2}$ intensity, calculating the Ru $3\text{d}_{3/2}$ intensity based on their known ratio, and deducting both of these intensities from the scan region. Note that for simplicity in the Ru case, the total C~1s intensity was assumed to come from graphene, as the deconvolution in FIG.\ \ref{fig:figure4}d revealed the bulk SiC signal to be negligible except when using the very highest ($h\nu>\SI{800}{\eV}$) photoexcitation energies. Both systems show a negative exponential decay in graphene intensity with increasing $h\nu$, as expected from Beer-Lambert's law for a system with finite signal detected from the surface layers. Based on these observations, we used a simple two-layer model to determine the thickness of the graphene layers, assuming these to be residing on top of underlying transition metal and SiC. For each photoexcitation energy, the signal from surface carbon $I_{s}$ was compared to the signal assigned to the underlying layers $I$ within the same scan region, according to
\begin{equation}\label{eq:layerEq}
    \frac{d}{\lambda} = \ln{\left[1+\frac{I\sigma_{s}}{I_{s}\sigma}\right]},
\end{equation}
where $\sigma_{s}$ and $\sigma$ are the photoexcitation cross-sections for the topmost and lowermost layers, respectively, and $d$ is the estimated thickness of the carbon overlayer. The total surface carbon signal of the gra/Fe/SiC system was thus compared to that of the the underlying bulk SiC, whereas for the gra/Ru/SiC system, the total C~1s signal was compared to that of the underlying Ru. FIG.\ \ref{fig:figure4}c shows that a linear relationship between the logarithm in Eqn.~\ref{eq:layerEq} and the inverse inelastic mean-free path is obtained for both materials systems, in good accordance with the assumed layer structure from the simplified two-layer model. The gradient from the best fit of each data set gives an estimate of the thickness $d$ of the graphene overlayers: \SI{1.1}{\nm} and \SI{2.6}{\nm} is obtained for the gra/Fe/SiC and gra/Ru/SiC systems, respectively. If one assumes that the graphene layers are weakly bound together by van der Waals forces with an interlayer spacing equal to that of graphite (\SI{0.355}{\nm})\cite{de2008strong}, then roughly 3-8 layers has been formed, regulated by the applied temperature and the thickness of the mediating metal deposited.

\section*{Conclusions}
The metal-mediated growth of epitaxial few-layer graphene on the surface of 6H-SiC(0001) treated with Fe or Ru has been investigated, at a temperature far lower than that required for graphene growth directly from the SiC crystal. Using surface diffraction and surface sensitive photoemission measurements, the onset of graphene formation was observed after short thermal treatments to \SI{450}{\celsius}, and the familiar electronic structure of graphene was confirmed for the newly formed species at \SI{600}{\celsius} using ARPES. Further annealing to higher temperatures 700-\SI{800}{\celsius} revealed the formation of additional graphene layers (3-8 layers). The placement of the graphene near the sample surface and its parallel orientation relative to the underlying growth substrate was established from depth sensitive photoemission and angle-dependent absorption spectroscopy measurements, respectively. The tunable depth sensitivity of the photoemission measurements was also used to confirm that the mediating metal agents, Fe and Ru, can be made to diffuse into the bulk SiC crystal with subsequent thermal treatments to higher temperatures.

From these investigations we have established a modified recipe for graphene production requiring a minimal number of processing steps. Controllable growth of high quality, few-layer graphene-on-semiconductor has been achieved, at industrially compatible temperatures that so far have not been available using other, more conventional growth techniques. The number of graphene layers formed is limited by the type and thickness of the transition metal film, and the temperatures to which the material system is subjected. The option to use either Fe or Ru interchangeably over a range of different temperatures allows  graphene with a tunable thickness to be formed. The possibility to diffuse the mediating metal agents into the substrate means graphene can either be supported on magnetic silicide layers or directly on SiC, as required. This makes metal-mediated graphene growth a realistic and promising candidate for realising graphene-based devices within the pre-existing framework of large-scale device processing.

\section*{Methods}
Single crystal samples of $n$-type 6H-\ce{SiC}(0001) (Tankeblue Inc.) were initially cleaned \emph{ex situ} using standard ``RCA'' chemical cleaning procedures to remove residual contaminants and native oxides \cite{kern1990evolution}. The samples were then subsequently loaded into the spectrometers and degassed \emph{in situ} at $\sim\SI{300}{\celsius}$ in ultrahigh vacuum (UHV) for durations of 6+ hours, before being rapidly annealed to 800-\SI{900}{\celsius} several times to remove any remaining oxide layers. The preparation of clean 6H-\ce{SiC}(0001) was confirmed using XPS and LEED.

Fe and Ru films with different thicknesses in the range 0.5-\SI{2.0}{\nm} were deposited on individual samples at constant rate using calibrated e-beam evaporation cells. The transition metal covered samples were then annealed for short durations at increasing temperatures ranging from $450$ to $\SI{700}{\celsius}$, to study the formation of transition metal (TM) silicides and the associated liberation of carbon atoms and $\text{sp}^2$ reconstruction at the surface. Subsequent annealing to $\sim\SI{800}{\celsius}$ for longer duration were performed on some samples to further study the thermally activated graphitization, and to investigate any associated changes to the concentration level of metal in the surface layers. Measurements were performed both prior to and following the thermal treatments to monitor the change with every experimental step. The formation of silicides and new carbon allotropes was confirmed through photon energy-dependent XPS measurements of the \ce{Si} 2p and \ce{C} 1s core levels, and where possible, NEXAFS measurements of the C~1s K-edge. The experiment was repeated and monitored using selective area low-energy electron diffraction ($\mu$-LEED) at every relevant preparation stage. Following the graphene growth, small-spot $\mu$-ARPES was also extracted from the diffractive plane of X-ray photoemission electron microscopy (XPEEM) measurements of the Fe/SiC system.

High-energy-resolution XPS measurements were performed at the MatLine and SXR endstations of the synchrotron radiation sources ASTRID (Aarhus) and the Australian Synchrotron (Melbourne, Victoria), respectively. At MatLine, core levels \ce{Fe} 3p, \ce{Si} 2p, \ce{C} 1s, \ce{Ru} 3d and \ce{O} 1s were measured with photon energies ranging from 120-\SI{600}{\eV} using a SCIENTA SES-200 analyser, with energy resolutions 100-\SI{850}{\meV}. The exact photon energy used, and therefore the binding energy of each core level spectrum, was calibrated using the core level signal generated by second order harmonic light from the same excitation energy as the principal peak. At SXR, the same core levels were recorded using excitation energies ranging from 100-\SI{1200}{\eV} using a Specs Phoibos 150 analyser, with an overall energy resolution in this range of 150-\SI{175}{\meV}. The binding energy of each peak was calibrated to the Fermi level position of TM/SiC substrate recorded over the corresponding photon energy range. All core level spectra were fitted using the Levenberg-Marquardt nonlinear least square method, using asymmetric pseudo-Voigt approximations to the Mahan functions of each chemical component as described by Schmid \emph{et al}\cite{schmid2014new}. For the core level backgrounds, an ``active'' approximation to the Shirley-Vegh-Salvi-Castle background\cite{herrera2014practical} was used: individual contributions to the background from each chemical component peak were modeled by error functions, and fitted dynamically together with all peaks and background components to simulate the full photoemission signal of the region. 

NEXAFS measurements were performed at the SXR endstation, in partial electron yield (PEY) mode and using linearly polarized light at incidence angles roughly 20-$100^\circ$ relative to the sample plane. For each measurement of the C~1s K-edge, a reference sample of HOPG, as well as a grounded gold mesh were placed upstream from the analysis chamber. During each scan, NEXAFS spectra for the reference sample and the drain current on the \ce{Au} mesh were recorded simultaneously using a small portion of the beam. Additionally, the absolute photon intensities coming in to the analysis chamber were subsequently measured over the same scan regions using a calibrated photodiode. The recorded NEXAFS signal was normalized to the drain current on the Au mesh (for details, see Ref. \onlinecite{watts2006methods}), and the spectral response of the photodiode was used as a secondary normalization to correct for any carbon contamination on the \ce{Au} mesh. Using the well known exciton resonance in HOPG occuring at \SI{291.65}{\eV}, the spectra of the HOPG reference that was measured in parallel with each scan was used to absolutely calibrate the energy scale of the data via a rigid shift. All energy and intensity flux corrections were performed using the Quick As NEXAFS Tool (QANT)\cite{QANT}.

Selective area diffraction ($\mu$-LEED) measurements were performed at the BL3.2Ub (PEEM) endstation at the Synchrotron Light Research Institute (SLRI), Muang Distrinct, Thailand, and at the UE49-PGM-SMART endstation at BESSYII, Helmholtz-Zentrum Berlin, Germany. Additionally,  bandstructure ($\mu$-ARPES) measurements were performed at UE49-PGM-SMART. Samples of gra/Fe/SiC and gra/Ru/SiC were prepared in a similar manner as described above. Clean SiC(0001) were confirmed by the lack of oxygen visible from dispersive plane XPS measurements, and at UE49-PGM-SMART also a $(\sqrt{3}\times\sqrt{3})\text{R}30^\circ$ silicate reconstruction of the surface. Metal was deposited using calibrated Fe and Ru sources and the thickness of the films were confirmed from XPEEM and low-energy electron microscopy (LEEM) measurements. All $\mu$-LEED measurements were performed using \SI{40}{\eV} excitation energy, except for the LEED pattern after graphene formation at \SI{620}{\celsius} (gra/Fe/SiC) and \SI{800}{\celsius} (gra/Ru/SiC): here, \SI{140}{\eV} and \SI{50}{\eV} were used, respectively, to reveal both the features native to graphene and the underlying SiC substrate. The ARPES of the graphene formed was recorded at room temperature, with excitation energy $h\nu=\SI{115}{\eV}$ and an overall energy resolution of \SI{500}{\meV}.

\section{Acknowledgements}
This work was partly supported by the Research Council of Norway through its Centres of Excellence funding scheme, project number 262633 ``QuSpin'', through project number 250555/O70 ``GraSeRad'', and through the Norwegian Micro- and Nano-Fabrication Facility, NorFab, project number 245963/F50. The SMART instrument was financially supported by the Federal German Ministry of Education and Research (BMBF) under the contract 05KS4WWB/4, as well as by the Max-Planck Society. Parts of this research was undertaken on the UE49-PGM-SMART beamline at BESSYII, the Synchrotron Light Research Institute (SLRI) in Thailand, and on the soft X-ray spectroscopy beamline at the Australian Synchrotron, part of ANSTO. We thank both the Helmholtz-Center Berlin for Materials and Energy (HZB), SLRI and ANSTO for the allocation of beamtime. S.P.C. would like to acknowledge the European Regional Development Fund (ERDF) and the Welsh European Funding Office (WEFO) for funding the 2nd Solar Photovoltaic Academic Research Consortium (SPARC II). B.P.R. acknowledges the EPSRC CDT in Diamond Science and Technology. L.d.S.C. is grateful for the funding through the Deutsche Forschungsgemeinschaft (DFG, German Research Foundation) under Germany's Excellence Strategy-EXC 2008-390540038 (UniSysCat). H.I.R., J.W.W. and S.P.C. would also like to thank Dr.\ Mark Edmonds and Dr.\ Andrew Evans for fruitful discussions.


%

\end{document}